\documentclass[prd,aps,nofootinbib,twocolumn,preprintnumbers]{revtex4}
\usepackage[utf8]{inputenc}
\usepackage[colorlinks=true,citecolor=blue,linkcolor=blue]{hyperref}
\usepackage[normalem]{ulem}
\usepackage{amsmath,amssymb, mathrsfs}
\usepackage{epsfig}
\usepackage{graphicx}               
\usepackage{url}
\usepackage{color}
\usepackage{slashed}
\usepackage{multirow}
\usepackage{placeins}
\usepackage[dvipsnames]{xcolor}
\usepackage{epstopdf}
\usepackage{soul}
\usepackage{tikz}
\usepackage[capitalise, english]{cleveref}
\usepackage{siunitx}
\usepackage{xspace}
\usepackage{booktabs}
\usetikzlibrary{trees}
\usetikzlibrary{decorations.pathmorphing}
\usetikzlibrary{decorations.markings}

\newcommand\myshade{80}
\colorlet{mylinkcolor}{ForestGreen}
\colorlet{mycitecolor}{Red}
\colorlet{myurlcolor}{violet}

\hypersetup{
  linkcolor  = mylinkcolor!\myshade!black,
  citecolor  = mycitecolor!\myshade!black,
  urlcolor   = myurlcolor!\myshade!black,
  colorlinks = true
}

\definecolor{jblue}{RGB}{20,50,100}
\definecolor{npurple}{RGB} {153, 51, 204}
\definecolor{wred}{RGB}{217,0,56}
\definecolor{white}{RGB}{255,255,255}

\definecolor{korange}{RGB}{235, 80,  43}
\definecolor{korange2}{RGB}{245, 100,  63}
\definecolor{kyelloworange}{RGB}{255, 210,  110}
\definecolor{kyelloworange2}{RGB}{240, 170,  90}
\definecolor{kred}{RGB}{204,  102, 153}
\definecolor{kpurple}{RGB}{153,  61, 190}
\definecolor{kpurplelight}{RGB}{213,  161, 230}

 \definecolor{tobycolour}{rgb}{.5,.0,.5}

\DeclareSIUnit\year{yr}
\DeclareSIUnit\pc{pc}
\DeclareSIUnit\ergs{ergs}
\DeclareSIUnit\msun{\ensuremath{M_\odot}}
\allowdisplaybreaks

\makeatletter
\providecommand*{\diff}%
  {\@ifnextchar^{\DIfF}{\DIfF^{}}}
\def\DIfF^#1{%
  \mathop{\mathrm{\mathstrut d}}%
    \nolimits^{#1}\gobblespace}
\def\gobblespace{%
  \futurelet\diffarg\opspace}
\def\opspace{%
  \let\DiffSpace\!%
  \ifx\diffarg(%
    \let\DiffSpace\relax
  \else
    \ifx\diffarg[%
      \let\DiffSpace\relax
    \else
        \ifx\diffarg\{%
        \let\DiffSpace\relax
      \fi\fi\fi\DiffSpace}

\usepackage{tikz,xcolor,hyperref}

\definecolor{lime}{HTML}{A6CE39}
\DeclareRobustCommand{\orcidicon}{\hspace{-1mm}
	\begin{tikzpicture}
	\draw[lime, fill=lime] (0,0) 
	circle [radius=0.16] 
	node[white] {{\fontfamily{qag}\selectfont \tiny \,ID}};
	\draw[white, fill=white] (-0.0525,0.095) 
	circle [radius=0.007];
	\end{tikzpicture}
	\hspace{-3mm}
}

\foreach \x in {A, ..., Z}{\expandafter\xdef\csname orcid\x\endcsname{\noexpand\href{https://orcid.org/\csname orcidauthor\x\endcsname}
			{\noexpand\orcidicon}}
}


\begin{document}

\title{Search for dark matter using sub-PeV $\boldsymbol{\gamma}$-rays observed by Tibet AS$_{\boldsymbol{\gamma}}$}

\author{Tarak Nath Maity\orcidA{}}
\email{tarak.maity.physics@gmail.com}
\affiliation{Centre for High Energy Physics, Indian Institute of Science, C.\,V.\,Raman Avenue, Bengaluru 560012, India}

\author{Akash Kumar Saha} 
\email{akashks@iisc.ac.in}
\affiliation{Centre for High Energy Physics, Indian Institute of Science, C.\,V.\,Raman Avenue, Bengaluru 560012, India}

\author{Abhishek Dubey} 
\email{abhishekd1@iisc.ac.in}
\affiliation{Centre for High Energy Physics, Indian Institute of Science, C.\,V.\,Raman Avenue, Bengaluru 560012, India}

\author{Ranjan Laha\orcidD{}} 
\email{ranjanlaha@iisc.ac.in}
\affiliation{Centre for High Energy Physics, Indian Institute of Science, C.\,V.\,Raman Avenue, Bengaluru 560012, India}

\date{\today}


\begin{abstract}
The discovery of diffuse sub-PeV gamma-rays by the Tibet AS$_\gamma$ Collaboration promises to revolutionize our understanding of the high-energy astrophysical universe.  It has been shown that these data broadly agree with prior theoretical expectations.  We study the impact of this discovery on a well-motivated new physics scenario: PeV-scale decaying dark matter (DM).  Considering a wide range of final states in DM decay, a number of DM density profiles, and numerous astrophysical background models, we find that these data provide the most stringent limit on DM lifetime for various Standard Model final states.  In particular, we find that the strongest constraints are derived for DM masses in between a few PeV to a few tens of PeV.  Near-future data of these high-energy gamma-rays can be used to discover PeV-scale decaying DM.
\end{abstract}

\maketitle

\section{Introduction}
\label{sec:introduction}

What is the {\it best} way to search for dark matter (DM) in a particular mass range?  Given the large number of astrophysical evidences of DM and a much larger list of well-motivated DM candidates, it is important to answer the question by considering various techniques and observables\,\cite{Aghanim:2018eyx, Strigari:2013iaa, Lisanti:2016jxe, Slatyer:2017sev, Lin:2019uvt, Feng:2010gw}.  Indirect detection of DM, in which one tries to detect the annihilating or decaying signature of DM via astrophysical measurements, is a promising way to discover various different DM candidates\,\cite{PerezdelosHeros:2020qyt}.  Because of the lack of knowledge about DM candidate masses, one must conduct the program of DM indirect detection using as many different observables as possible.  

One such promising observable is high-energy gamma-rays (energies $E \gtrsim$ 100 TeV).  These high-energy gamma-rays may be produced by hadronic or leptonic processes\,\cite{hayakawa1952propagation, Kelner:2006tc, kappes2007potential, Gupta:2011aw, Lipari:2018gzn}.  They can either be produced near astrophysical objects or can be produced by cosmic-ray (CR) interaction with interstellar gas.  Since the mean free path of photons with energies between $\sim$ 100 TeV and $\sim$ 10$^{19}$ eV is $\lesssim$ few Mpc, a detection of gamma-rays in this energy range primarily probes the high-energy astrophysics of the Milky Way (MW)\,\cite{Moskalenko:2005ng, Venters:2010bq, Vernetto:2016alq, DeAngelis:2013jna, ruffini2016cosmic}.  The search for these gamma-rays will also indicate whether the Milky Way galaxy is a ``PeVatron."

\begin{figure}
\centering
\includegraphics[width=\columnwidth]{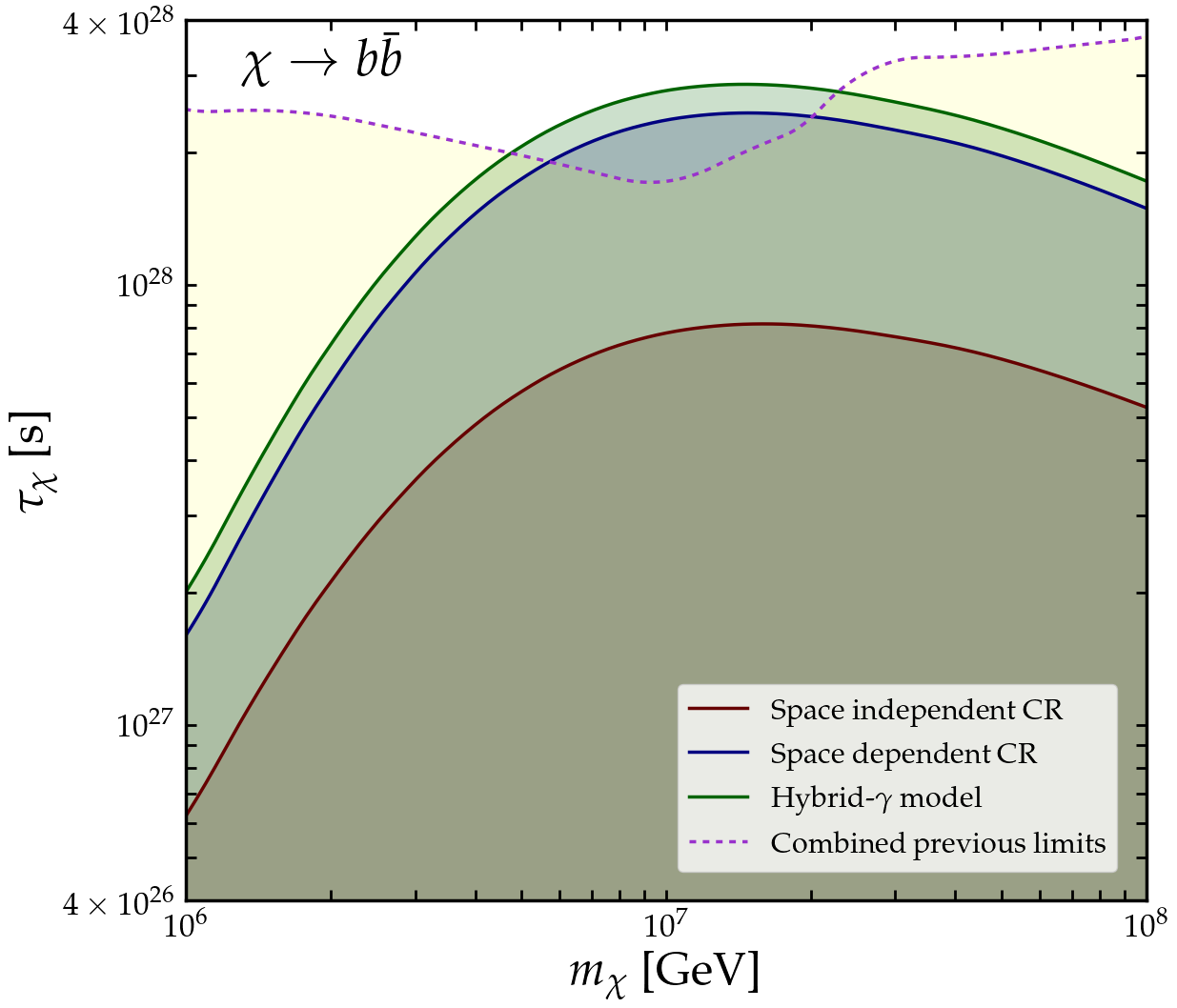}
\caption{Upper limit on the lifetime of DM, $\tau_\chi$ (s), as a function of its mass $m_\chi$ (GeV).  Our constraints are shown by the solid lines.  The blue, red, and orange lines represent the upper limits when the astrophysical background follows the space-independent cosmic-ray prediction, space-dependent cosmic-ray prediction, and prediction from Fang {\it et al}., hybrid-$\gamma$ model as given in Refs.\,\cite{Amenomori:2021gmk} and \cite{Fang:2021ylv}.  The dashed line shows the best upper limit on $\tau_\chi$ from previous studies\,\cite{Cohen:2016uyg, Blanco:2018esa, Bhattacharya:2019ucd, Ishiwata:2019aet, Chianese:2019kyl}.  We assume the DM decay channel $\chi \rightarrow b \bar{b}$ for this plot.  We assume that the DM density profile follows the NFW form in this plot and consider the 25$^\circ$ $<$ $\ell$ $<$ 100$^\circ$, $|b|$ $<$ 5$^\circ$ sky region.}
  \label{fig:DM to b+b-bar} 
\end{figure}

Recently, Tibet AS$_\gamma$ Collaboration discovered the first sub-PeV diffuse gamma-rays from the MW Galactic disk, opening up a new chapter in gamma-ray astrophysics\,\cite{Amenomori:2021gmk} (earlier upper limits in a similar energy range exist from the CASA-MIA and Kascade collaborations\,\cite{Chantell:1997gs, Borione:1997fy, Apel:2017ocm}).  These gamma-rays, which have energies between $\sim$ 100 TeV and 1 PeV, were detected from two sky regions: (a) 25$^\circ$ $<$ $\ell$ $<$ 100$^\circ$, $|b|$ $<$ 5$^\circ$ and (b) 50$^\circ$ $<$ $\ell$ $<$ 200$^\circ$, $|b|$ $<$ 5$^\circ$, where $b$ and $\ell$ denote the Galactic latitude and longitude, respectively.  Studying the angular morphology and the energy spectrum of these photons, the Tibet AS$_\gamma$ Collaboration concluded that these high-energy gamma-rays are produced via CR interaction with interstellar gas.  The collaboration also demonstrated that these observations follow prior theoretical expectations.  After the publication of this discovery, a number of works revisited the high-energy diffuse gamma-ray production in the MW due to hadronic and leptonic processes\,\cite{Dzhatdoev:2021xjh, Qiao:2021iua, Liu:2021lxk, Fang:2021ylv, Koldobskiy:2021cxt}.  These works confirmed that the Tibet AS$_\gamma$ data can be reproduced using viable astrophysical parameters.

The detection of these diffuse gamma-rays also opens a new window into the study of new physics.  One of the well-motivated models of new physics which can potentially be discovered via these photons is decaying DM\,\cite{Murase:2015gea, Esmaili:2015xpa, Esmaili:2014rma, Kalashev:2016cre, Kachelriess:2018rty, Ishiwata:2019aet}.  Among the simplest models are those in which the DM decays into two SM particles.  These final-state particles produce gamma-rays due to decays and electroweak processes.  The current best constraints on heavy DM (masses in between $\sim$ PeV and $\sim$ 10 PeV) come from the observations of the diffuse astrophysical neutrino flux by IceCube and the diffuse gamma-ray background by  Fermi-LAT\,\cite{Cohen:2016uyg, Bhattacharya:2019ucd, Ishiwata:2019aet, Chianese:2019kyl, Kachelriess:2018rty}.  Neutrinos can be produced from various SM final states via decay and electroweak processes, and one can search for these neutrinos in the high-energy astrophysical neutrino dataset.  Heavy DM can produce high-energy photons which can get attenuated through interaction with the low-energy photons present in the Universe.  This attenuation process produces $e^\pm$ pairs which upscatter low-energy photons in the Universe.  These upscattered photons can be detected by Fermi-LAT.  These two techniques can be used to produce some of the strongest constraints on PeV-scale DM.

In this work, we demonstrate that this new discovery of sub-PeV gamma-rays in the Galactic disk region can be used to probe new regions of PeV-scale decaying DM.  Given the initial dataset and its broad agreement with theoretical predictions, the current dataset does not give any hint of PeV-scale decaying DM signal.  However, we show that this dataset can produce the most stringent constraints on the lifetime of PeV-scale decaying DM.  In Fig.\,\ref{fig:DM to b+b-bar}, we show the upper limits on DM lifetime, $\tau_\chi$, that we obtain for the $\chi \rightarrow b \bar{b}$ channel, where $\chi$ denotes the DM particle.  The previous best limits for this channel are combined and shown by the dashed line\,\cite{Cohen:2016uyg, Bhattacharya:2019ucd, Ishiwata:2019aet, Chianese:2019kyl}.  We take into account various different astrophysical background models to derive our limits.  The three kinds of backgrounds that we consider are: the space independent (SI) and space dependent (SD) CR models from Refs.\,\cite{Lipari:2018gzn,Amenomori:2021gmk} and the hybrid-$\gamma$ model from Ref.\,\cite{Fang:2021ylv}. For both the sky regions a and b, without including a DM contribution, the SI CR model yields a relatively poor fit to the data, and the constraint obtained using this background is the weakest. For region a, much better fits to the data are obtained with the SD CR and hybrid-$\gamma$ model (without DM contribution), and these two limits are most representative of what can be extracted from the dataset.  The data from the sky region 25$^\circ$ $<$ $\ell$ $<$ 100$^\circ$, $|b|$ $<$ 5$^\circ$ and NFW DM profile are used to obtain our results.\footnote{Note that for this region of the observation,  with Space dependent CR and Hybrid-$\gamma$ Model, inclusion of DM flux does not improve these fits.} Using other astrophysical models responsible for the detected gamma-rays\,\cite{Dzhatdoev:2021xjh, Qiao:2021iua, Liu:2021lxk, Fang:2021ylv, Koldobskiy:2021cxt} result in a limit which is very similar to that shown in this figure.

Considering a wide variety of astrophysical backgrounds and various two-body SM final states, we find that our constraints are better than various previous constraints for most of the channels.  This is a very promising result; it demonstrates that near-future data on sub-PeV gamma-rays from regions of the Galaxy which contain a higher DM density have the potential to discover DM.  Our results imply that a better understanding of the production mechanism, energy spectrum, and angular spectrum of these photons, along with a near-future larger dataset of sub-PeV gamma-rays will allow one to show that this is the most superior technique of searching PeV-scale decaying DM.

\section{Formalism}
\label{sec:formalism}

PeV-scale decaying DM acts as cold DM and thus its mass density profiles inside DM halos follows the Navarro Frenk White (NFW) or Einasto type\,\cite{navarro1997universal, einasto1989galactic, mcmillan2011mass, Chacon:2020blu, Vogelsberger:2019ynw}.  For the MW galaxy, baryonic processes can produce a cored DM profile and various measurements exist of the MW DM profile\,\cite{mollitor2015baryonic, chan2015impact, Benito:2020lgu}.  For concreteness, we assume that the DM density profile in the MW galaxy can be parametrized by the form $\rho_\chi^{\alpha \beta \gamma} (r) \,=\, \rho_\odot \, \left(\dfrac{r}{r_\odot} \right)^{- \gamma} \, \left(\dfrac{1 \,+\, (r_\odot/r_s)^\alpha}{1 \,+\, (r/r_s)^\alpha} \right)^{(\beta - \gamma)/\alpha}$, where $\rho_\odot$ = 0.4 GeV cm$^{-3}$ is the local DM density, $r$ is the distance from the Galactic Center, $r_\odot$ = 8.3 kpc is the distance to the MW center from the Solar position\,\cite{Abuter:2020dou}, and $r_s$ denotes the scale radius of the DM halo\,\cite{Ng:2013xha}.  The NFW profile is represented by $\alpha$ = 1, $\beta$ = 3, $\gamma$ = 1, and $r_s$ = 20 kpc.  The cored profile is represented by $\alpha$ = 2, $\beta$ = 2, and $\gamma$ = 0.  A representative value of $r_s$ for the cored profile is 3.5 kpc.  The Einasto DM profile can be represented as $\rho_\chi^{\rm Ein} (r) \,=\, \rho_\odot \, {\rm exp}\,\left(- \dfrac{2}{0.17}\,\dfrac{r^{0.17}\,-\, (8.3 \, {\rm kpc})^{0.17}}{(20\,{\rm kpc})^{0.17}} \right)$. 

Various particle physics models of DM have been proposed in the literature where the DM mass is in $\sim$ PeV range\,\cite{Feldstein:2013kka, Harigaya:2013pla, Harigaya:2014waa, Bhattacharya:2014yha, Rott:2014kfa, Dudas:2014bca, Daikoku:2015vsa, Kopp:2015bfa, Aisati:2015vma, Anchordoqui:2015lqa, Boucenna:2015tra, Ko:2015nma, Aisati:2015ova, Dev:2016qbd, Roland:2015yoa, Dev:2016uxj, ReFiorentin:2016rzn, DiBari:2016guw, Chianese:2016smc, Cata:2016epa, Bhattacharya:2016tma, Hiroshima:2017hmy, Borah:2017xgm, Chakravarty:2017hcy, Chianese:2017nwe, Dhuria:2017ihq, Sui:2018bbh, Lambiase:2018yql, Xu:2018qnd, Chianese:2018ijk, Kim:2019udq, Jaeckel:2020oet, Garcia:2020hyo, Azatov:2021ifm, Jaeckel:2021gah}.  The final-state SM particles produced in DM annihilation or decay is dependent on the underlying DM model.  In this work, we will be agnostic of the underlying particle physics model and take various two-body SM final states.  Although, our choices of SM final states are clearly not complete, we estimate that our limits will probe well-motivated parameter space of various possible particle physics models.  For simplicity, we only consider DM decay, although the same dataset can be used to constrain annihilating DM, too.  We consider charged and neutral leptons, gauge bosons, quarks, and Higgs bosons in the final state of DM decay.  These final-state particles can hadronize (if they are quarks) or decay (if they are unstable) to produce high-energy gamma-rays.  Even a stable SM particle can produce high-energy gamma-rays via electroweak bremsstrahlung.  There have been a number of studies on electroweak bremsstrahlung in recent years\,\cite{Bell:2008ey, Kachelriess:2009zy, Cirelli:2010xx, Ciafaloni:2010ti, Bauer:2020jay}, and we take the photon spectrum from {\tt HDMSpectra}\,\cite{Bauer:2020jay}. In our DM mass range of interest, the results of {\tt HDMSpectra} agree reasonably well with {\tt PPPC4DMID}\,\cite{Cirelli:2010xx}, and the differences between the final limits obtained with these spectra are smaller than 10\%.  These photons, emerging directly from the DM decay due to various SM final states, are also called the prompt photons.

The high-energy photons produced via electroweak bremsstrahlung and SM particle decays do not travel unimpeded from its production point to the observer.  These high-energy gamma-rays can interact with the low-energy photons, $\gamma \,+\, \gamma \rightarrow e^+ \,+\, e^-$, and get attenuated.  The various low-energy gamma-rays consist of cosmic microwave background (CMB), starlight (SL), infrared (IR), and extra-Galactic background light (EBL).  We have used the parametrization of these low-energy photons from Ref.\,\cite{Vernetto:2016alq} and used the standard formulas to calculate the attenuation factor\,\cite{Vernetto:2016alq}.  Because of the much smaller density of the EBL at all relevant frequencies, we neglect its impact in this work.  The impact of the SL + IR is most strongly realized when a high-energy photon passes through the Galactic Center, where the attenuation of the flux can be $\gtrsim$ 10\%, depending on the energy of the photon.  The regions of interest considered in this work do not include the Galactic Center, and the attenuation of the high-energy photons on the SL + IR is very marginal, $\sim$ $\mathcal{O}$(1\%), in our work.  Using other parametrizations of the SL + IR density will produce a similar marginal effect on the attenuation factor for high-energy gamma-rays.  The mean free path of photons with energy $\sim$ 500 TeV on the CMB is $\sim$ 20 kpc.    Because of the Galactic scale mean free paths of the photons detected by Tibet AS$_\gamma$ (with energies in between $\sim$ 100 TeV and $\sim$ 1 PeV), we only consider the DM decay in the MW.

Combining all this information, we write the prompt photon flux from DM decay as 
\begin{eqnarray}
\frac{d^2 \phi_{\gamma}}{d E_{\gamma} d\Omega}(E_{\gamma})& =  \dfrac{1}{\Delta \Omega}\,\int_{\Delta \Omega} d\Omega\frac{1}{4 \pi m_{\chi} \tau_{\chi}} \frac{dN_{\gamma}}{dE_{\gamma}}(E_{\gamma}) \nonumber \\
&\int_{0}^{s_{\rm max}} \rho_{\chi}(s,b,l)\,e^{-\tau_{\gamma \gamma}(E_{\gamma},s, b, \ell)} ds,
\label{eq: photon flux DM decay}
\end{eqnarray}
where $\Delta \Omega$ denotes the total angular size of the region of observation, $m_\chi$ denotes the DM mass, $\tau_\chi$ denotes the DM lifetime, $dN_\gamma/dE_\gamma$ denote the prompt photon emission spectrum, $E_\gamma$ denote the energy of the prompt photons, $\rho_\chi$ denote a generic DM density profile, $\tau_{\gamma \gamma}$ denote the high-energy photon attenuation factor, and $s$ denotes the line-of-sight distance.  The relation between $s$ and $r$ depends on $b$ and $\ell$: $r(s,\,b,\,\ell) \,=\, \sqrt{s^2 \,+\, r_\odot^2 \,-\, 2 \, s \, r_\odot \, {\rm cos}\, b \, {\rm cos}\,\ell}$. The differential angular window is written as d$\Omega$ = d\,(sin\,$b$)\,d$\ell$.

In addition to prompt photons, DM decays also produce high-energy electrons and positrons.  These high-energy $e^\pm$ lose energy during propagation.  One of the main energy loss channels is inverse Compton (IC), where these high-energy $e^\pm$ scatter up low-energy Galactic photons to high energy.  We study this IC channel following Ref.\,\cite{Cirelli:2009vg}.

Taking the measurements in the three energy bins provided by the Tibet AS$_\gamma$ Collaboration, we can write the $\chi^2$ estimator as
\begin{eqnarray}
\hat{\chi}^2 \,=\, \sum \left({\rm data} \,-\, {\rm astro.\,model} \,-\, {\rm NP\,model}\right)^2/\sigma^2\,,\phantom{x}
\label{eq: chi-squraed}
\end{eqnarray}
where astro.\,model indicates the flux due to some theoretical astrophysics model in this energy range, NP model indicates the flux due to new physics, and $\sigma$ denotes the error bars in the data.  We assume various theoretical astrophysics models advocated in recent literature.
\begin{figure}
\centering
\includegraphics[width=\columnwidth]{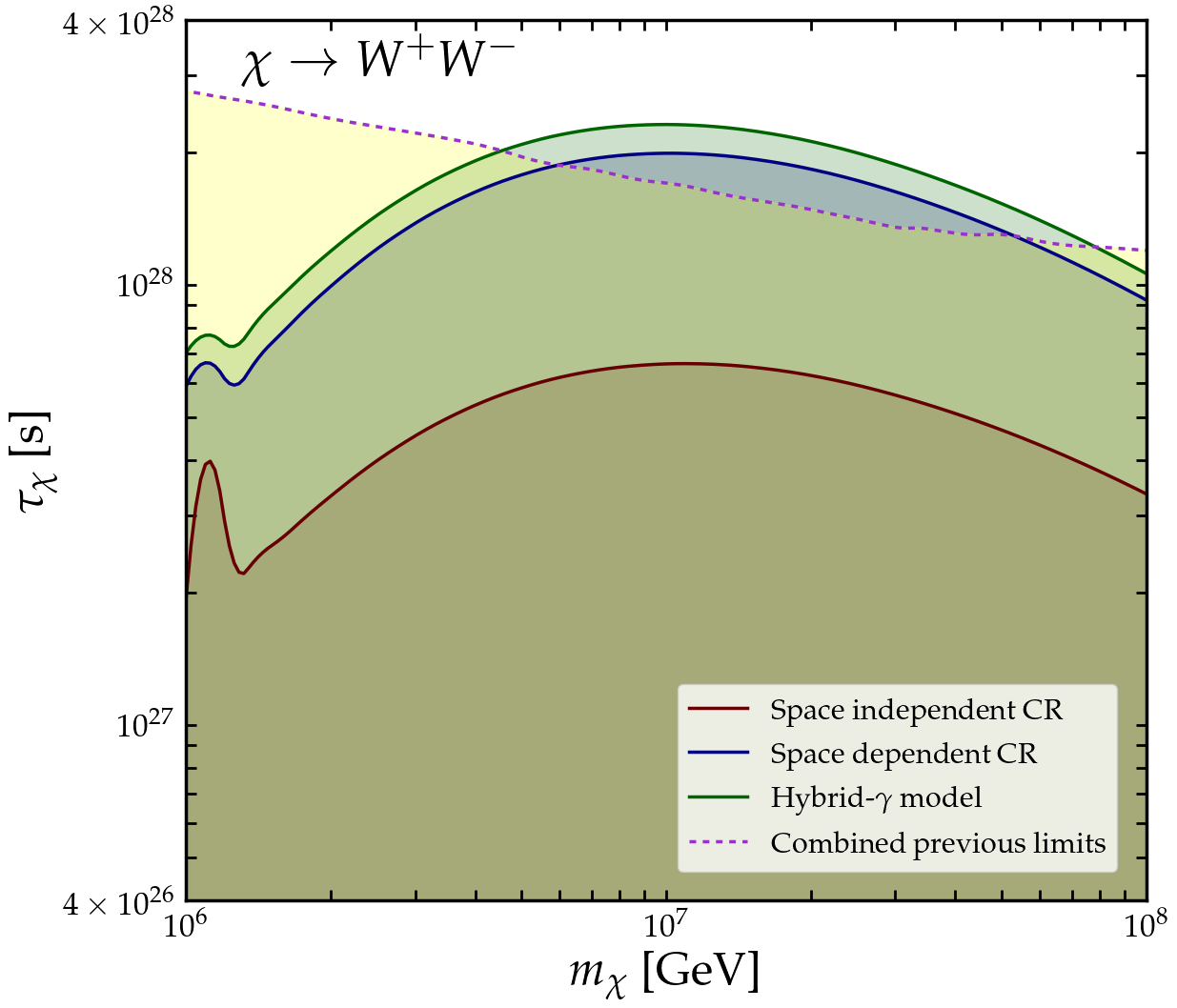}
\caption{Upper limit on the lifetime of DM, $\tau_\chi$ (s), as a function of its mass $m_\chi$ (GeV).  We assume the DM decay channel $\chi \rightarrow W^+ W^-$ for this plot.  The linestyles and the assumptions are same as those in Fig.\,\ref{fig:DM to b+b-bar}.}
\label{fig:DM to W+W-} 
\end{figure}
The NP model follows from Eq.\,\eqref{eq: photon flux DM decay}. In sky region 25$^\circ$ $<$ $\ell$ $<$ 100$^\circ$, $|b|$ $<$ 5$^\circ$, among  the considered theoretical astrophysical models, for the SD CR and hybrid-$\gamma$ models the measured data are reproduced and the minimum value of the $\hat{\chi}^2$( $\hat{\chi}^2_{\rm min}$) is obtained when the decaying DM contribution is absent. This is not the case for SI CR model  (one can observe a small excess in the highest energy bin considering the SD CR model for the sky region 50$^\circ$ $<$ $\ell$ $<$ 200$^\circ$, $|b|$ $<$ 5$^\circ$ and for both the sky regions with SI CR model. Since the observed excesses are $\sim 1 \sigma$ for the mentioned models, we do not consider this excess as an indication of a DM signature).  After calculating the value of $\hat{\chi}^2_{\rm min}$, we obtain the 95\% C.L.\,upper limit from the equation $\hat{\chi}^2 \,-\, \hat{\chi}^2_{\rm min} \approx$ 2.71 using Ref.\,\cite{Cowan:2010js,Lamperstorfer:2015cfg}.

\section{Results}
\label{sec:results}

The 95\% C.L.\,upper limits on $\tau_\chi$ for $\chi \rightarrow b \bar{b}$ and $\chi \rightarrow W^+ W^-$ are shown in Figs.\,\ref{fig:DM to b+b-bar} and \,\ref{fig:DM to W+W-}, respectively.  For $\chi \rightarrow b \bar{b}$, we see that this dataset produces the most stringent constraint for masses $\sim$ 6 PeV to $\sim$ 15 PeV.  Since the region of interest (25$^\circ$ $<$ $\ell$ $<$ 100$^\circ$) is far away from the Galactic Center, the constraint is almost independent of the DM profile.  Because of the lower energy density of the SL and IR in the region of interest, we find that the attenuation of the high-energy photons on these low-energy photon backgrounds is small.  Similarly due to the lower density of these low-energy photons, the IC contribution to our constraints is small.  The largest uncertainty in our derived limits comes from the unknowns in the theoretical prediction of this Galactic CR induced photon flux. Various hadronic and leptonic models have been proposed in the literature to explain this astrophysical flux, and a near future deeper understanding of the theory predictions will help sharpen the constraints.  We find that the constraints from the sky region 50$^\circ$ $<$ $\ell$ $<$ 200$^\circ$, $|b|$ $<$ 5$^\circ$ are weaker and hence we do not show them\footnote{We note that taking SI CR as the background model this region provides a slightly stronger ($\sim 6\%$) bound in DM mass range $10^8 \,{\rm GeV} \lesssim m_{\chi} \lesssim 10^9 {\rm GeV}$ as compared to the same arising from sky region 25$^\circ$ $<$ $\ell$ $<$ 100$^\circ$, $|b|$ $<$ 5$^\circ$. While for other considered DM parameter space region 25$^\circ$ $<$ $\ell$ $<$ 100$^\circ$, $|b|$ $<$ 5$^\circ$ provides better ($\sim \mathcal{O}(10)\%$) constraints.}.

When considering the DM decay channel $\chi \rightarrow W^+ W^-$, we find that our work results in the strongest constraint in the mass range $\sim $ 9 PeV to $\sim$ 30 PeV.  The previous leading constraint in this channel was provided in Refs.\,\cite{Cohen:2016uyg, Blanco:2018esa, Chianese:2019kyl}, and these are shown until $m_\chi$ = $10^3$ PeV.  Constraints at higher masses are derived in Ref.\,\cite{ Kachelriess:2018rty} from KASCADE-Grande\,\cite{Apel:2017ocm}, PAO\,\cite{Aab:2016agp}, TA\,\cite{Abbasi:2018ywn}, or IceCube data\,\cite{Aartsen:2016ngq}; however, in this DM mass range, we obtained a subdominant limit from Tibet AS$_{\gamma}$, which thus has not been displayed.

In the supplemental material, we display further constraints on $\tau_\chi$ for various other final-state SM particles.  We observe that our derived constraints are the most stringent for a variety of final states: $\chi \rightarrow u \bar{u}$, $\chi \rightarrow d \bar{d}$, $\chi \rightarrow s \bar{s}$, $\chi \rightarrow c \bar{c}$, $\chi \rightarrow t \bar{t}$, $\chi \rightarrow g \bar{g}$, $\chi \rightarrow Z \bar{Z}$ and $\chi \rightarrow h h$. For $\chi \rightarrow \tau^+ \tau^-$ and $\chi \rightarrow \nu_\tau \bar{\nu}_\tau$ our results are superior to those obtained from Fermi-LAT data, and they are equally stringent as those obtained from IceCube data in some region of the parameter space.  For $\chi \rightarrow \nu_e \bar{\nu}_e$,  $\chi \rightarrow e^+ e^-$, $\chi \rightarrow \nu_\mu \bar{\nu}_\mu$, and $\chi \rightarrow \mu^+ \mu^-$, our derived limits are weaker than the previous limits.

\section{Conclusions}
\label{sec:conclusions}

It has been anticipated for decades that high-energy cosmic-ray interaction with interstellar gas will produce high-energy gamma-rays and this will give us a unique insight into the high-energy astrophysics of the MW.  Recently, the Tibet AS$_\gamma$ discovered a diffuse flux of these photons from the Galactic disk region.  The data match well with prior theoretical expectations.  We study the impact on PeV-scale decaying DM, focusing on the data obtained from the region 25$^\circ$ $<$ $\ell$ $<$ 100$^\circ$, $|b|$ $<$ 5$^\circ$.  Considering various SM final states and astrophysical production models of sub-PeV gamma-rays, we find that these data reveal the strongest constraint on decaying PeV-scale DM for most of the two-body final states.  In particular, the strongest constraints are obtained when the SM final states are quarks, gauge bosons, Higgs boson, $\tau^+ \tau^-$,  and $ \nu_\tau \bar{\nu}_\tau$.  For decay to other two-body SM final states that we have considered, our constraints are weaker than those obtained via IceCube observations.  Our constraints are robust with respect to variations in the DM density profile.  These results open up a new way to probe this class of DM models.  Although we have focused on decaying DM, these data can be used to  probe heavy annihilating DM, too, which we leave for future work.  These classes of DM models can also be well probed via various other observations\,\cite{Beacom:2006tt, Dasgupta:2012bd, Arguelles:2019ouk, Dekker:2019gpe, Ng:2020ghe, Chianese:2021htv, Ando:2021fhj}.  In addition, a near-future dataset from the Tibet AS$_\gamma$ Collaboration may help us discover the DM particle identity.

 \emph{Acknowledgments --} We  thank  Oscar Macias, Kenny C.\,Y.\,Ng, and Anupam Ray for discussions.  We also thank the referees for pointing out a bug in our calculation, which led to a $\sim$ $\mathcal{O}$(10\%) change in our results.  T.N.M. acknowledges the IoE-IISc fellowship program for financial assistance.
 
 \emph{Note added --- }  Reference \cite{Esmaili:2021yaw} appeared on arXiv recently. While Ref.\,\cite{Esmaili:2021yaw} has presented its DM constraints by considering only the space independent CR model. We have considered three representative sets of CR models to present our results, in order to encompass the underlying theoretical uncertainty in predicting the Galactic gamma-ray flux.  We have also considered a wider variety of two-body SM final states compared to Ref.\,\cite{Esmaili:2021yaw}. With the SI background CR model, the difference between our and Ref.\,\cite{Esmaili:2021yaw}'s results is $\sim$ $\mathcal{O}$(10\%) for spectral analysis.

All the first three authors contributed equally to this work.

\newpage
\bibliographystyle{JHEP}
\bibliography{ref.bib}

\clearpage
\newpage
\maketitle
\onecolumngrid
\begin{center}
\textbf{\large Supplementary Material} \\

\vspace{0.05in}

\textbf{\large Search for dark matter using sub-PeV $\boldsymbol{\gamma}$-rays observed by Tibet AS$_{\boldsymbol{\gamma}}$} \\

\vspace{0.07in}

{Tarak Nath Maity, Akash Kumar Saha, Abhishek Dubey, and  Ranjan Laha}

\end{center}

In this Supplementary Material we present limits on decaying DM for other two body final states and its comparison with previous limits.
\vspace{-0.8cm}
\section*{Other final states} 
While in the main text we have presented our results for two scenarios i.e., DM decaying to $b \, \bar{b}$ and $W^{+}\, W^{-}$.  In this section, we present the constraints on DM lifetime for several other final states. For each plot, the combined previous constraints from Refs.\,\cite{Cohen:2016uyg, Kachelriess:2018rty, Blanco:2018esa, Bhattacharya:2019ucd, Chianese:2019kyl} are also displayed.

\begin{figure*}[!h]
\begin{center}
	\includegraphics[angle=0.0,width=0.265\textwidth]{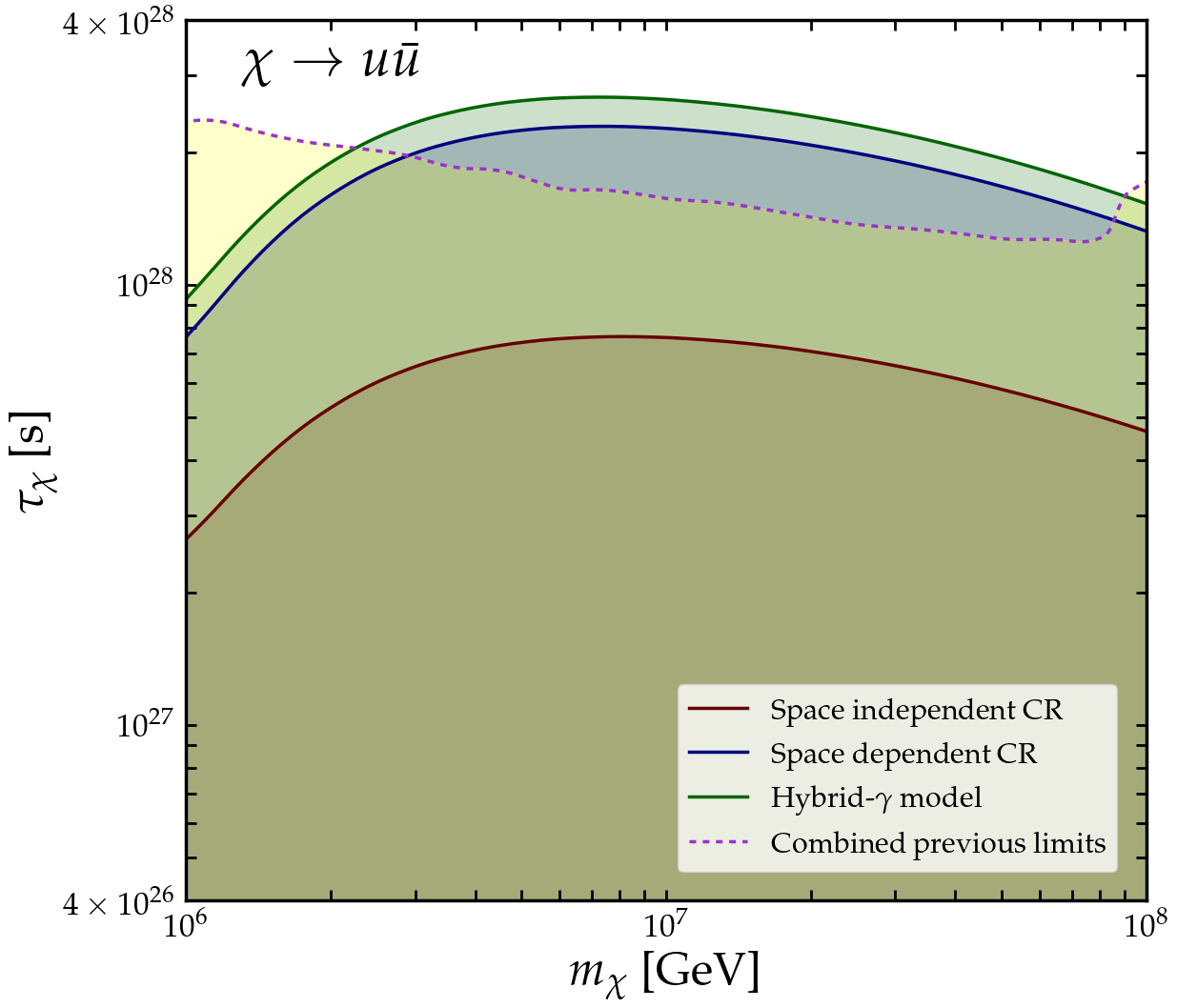}~~
	\includegraphics[angle=0.0,width=0.265\textwidth]{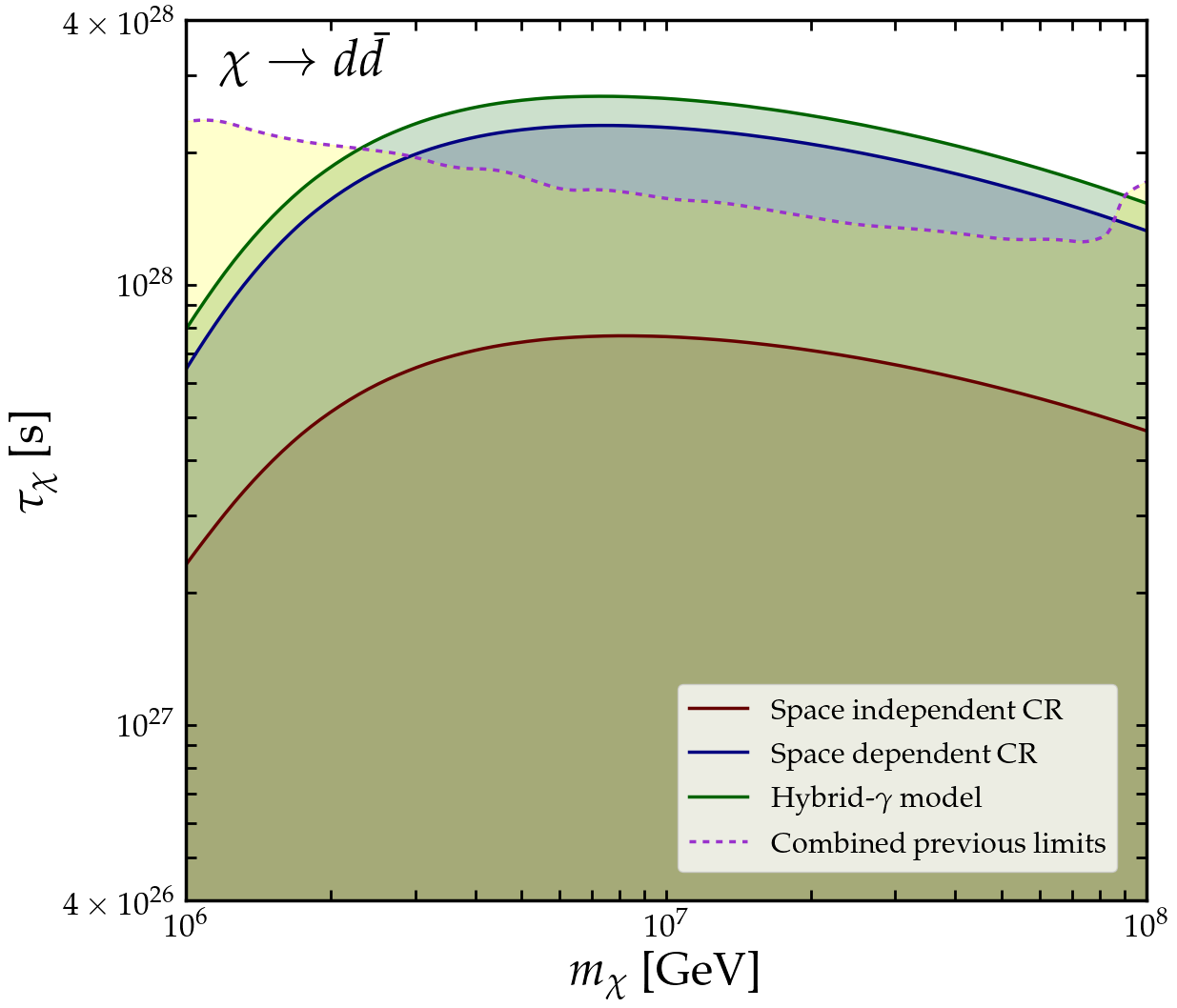}~~
	\includegraphics[angle=0.0,width=0.265\textwidth]{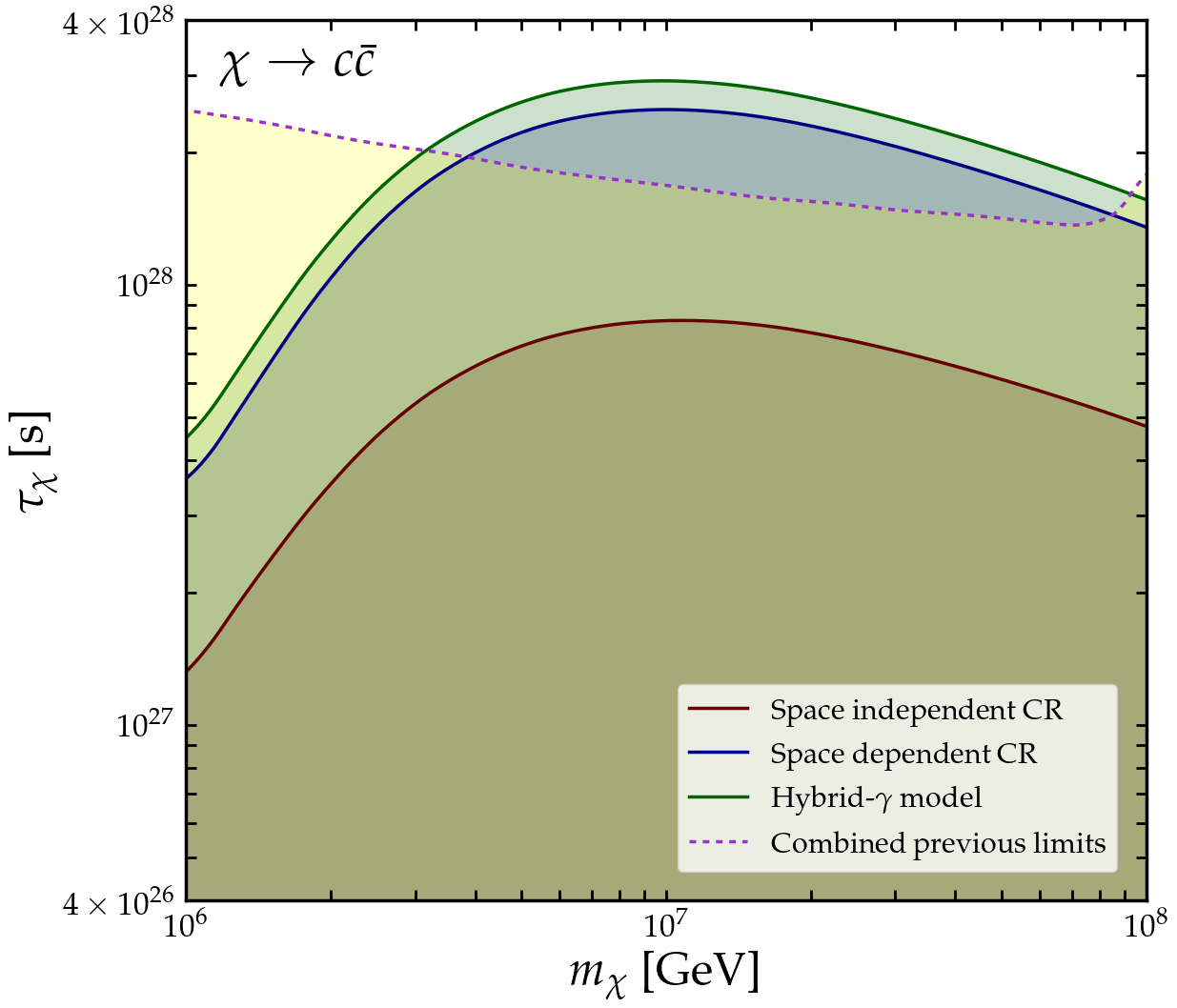}~~\\
	\includegraphics[angle=0.0,width=0.265\textwidth]{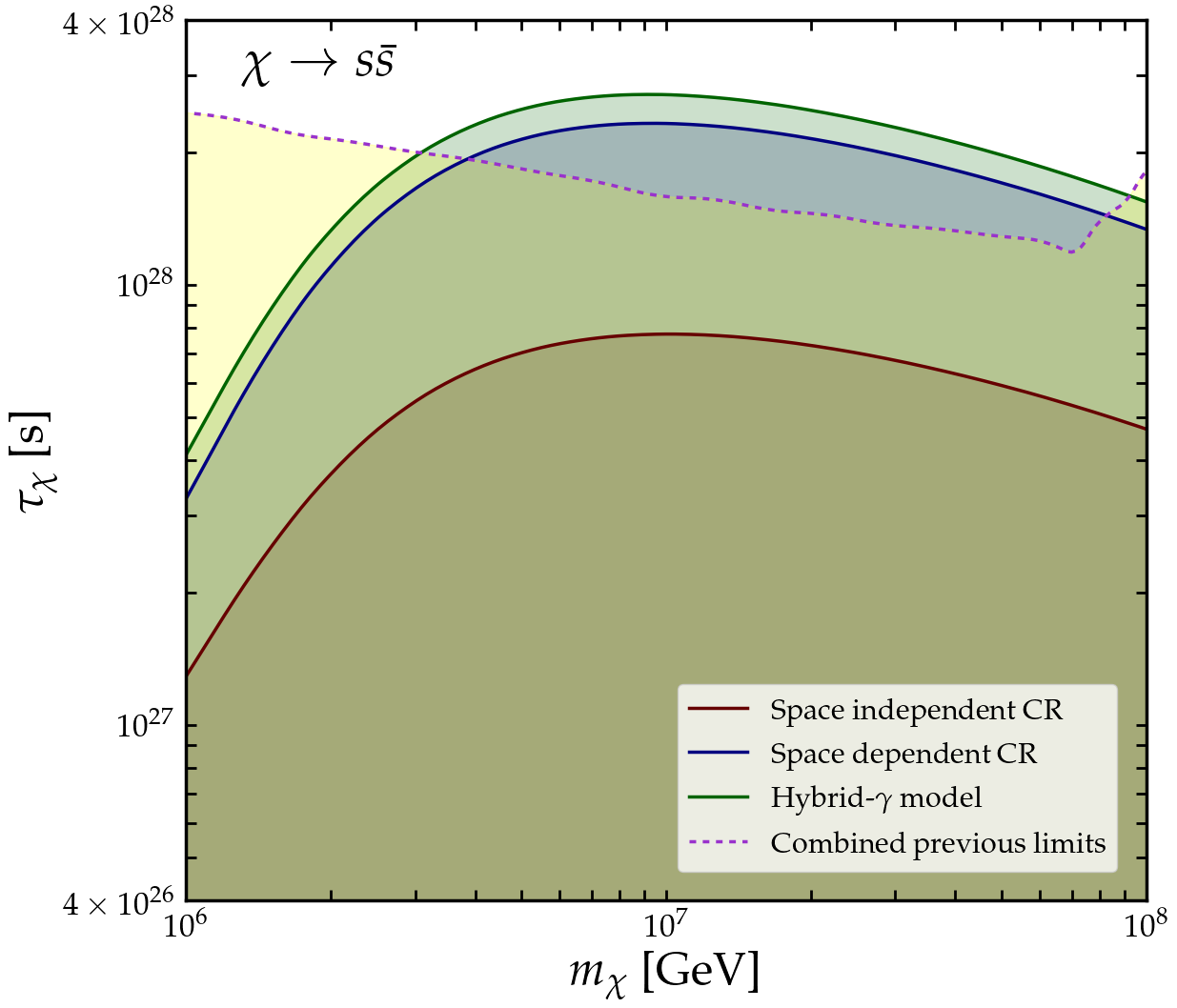}~~
	\includegraphics[angle=0.0,width=0.265\textwidth]{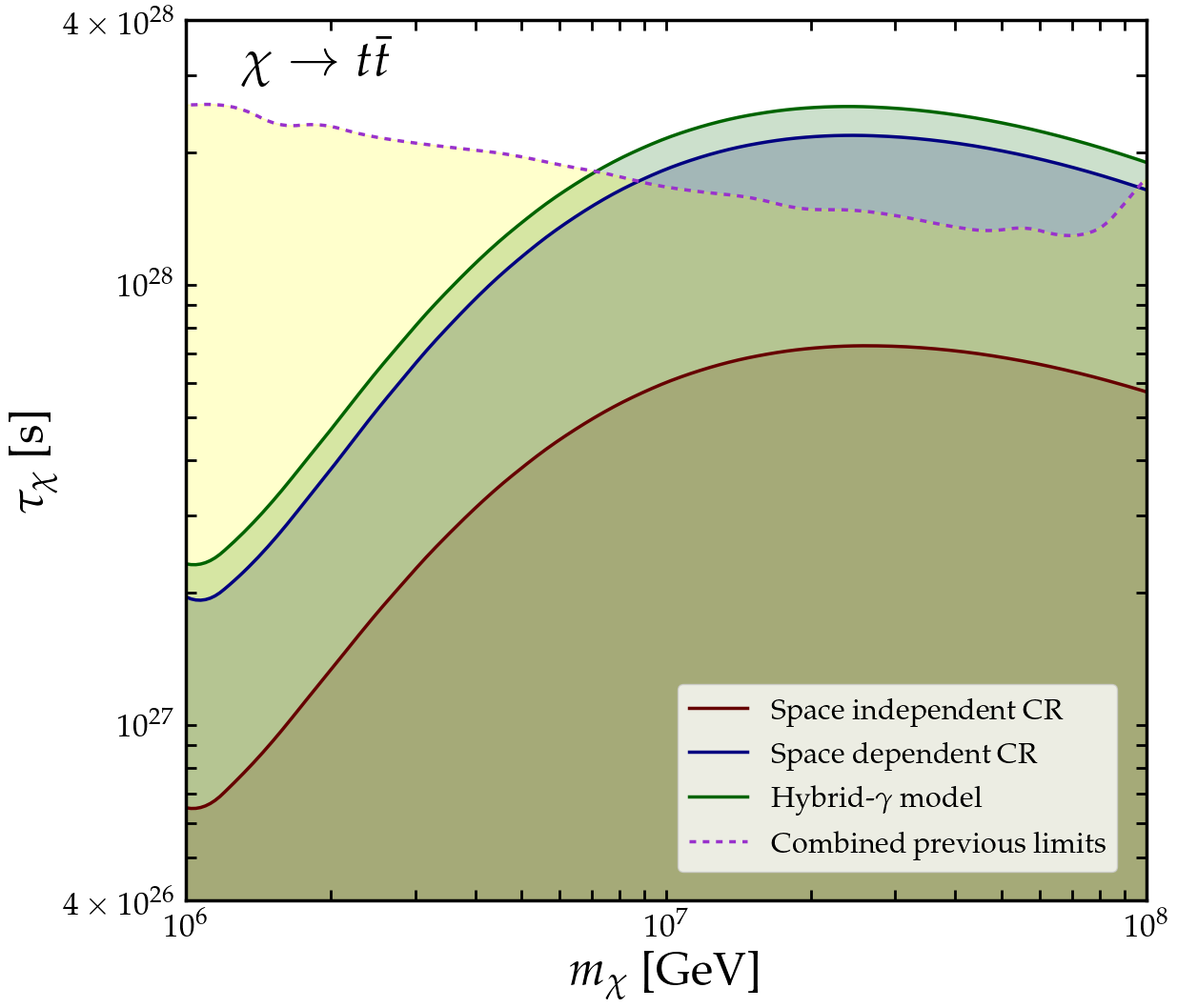}~~
	\includegraphics[angle=0.0,width=0.265\textwidth]{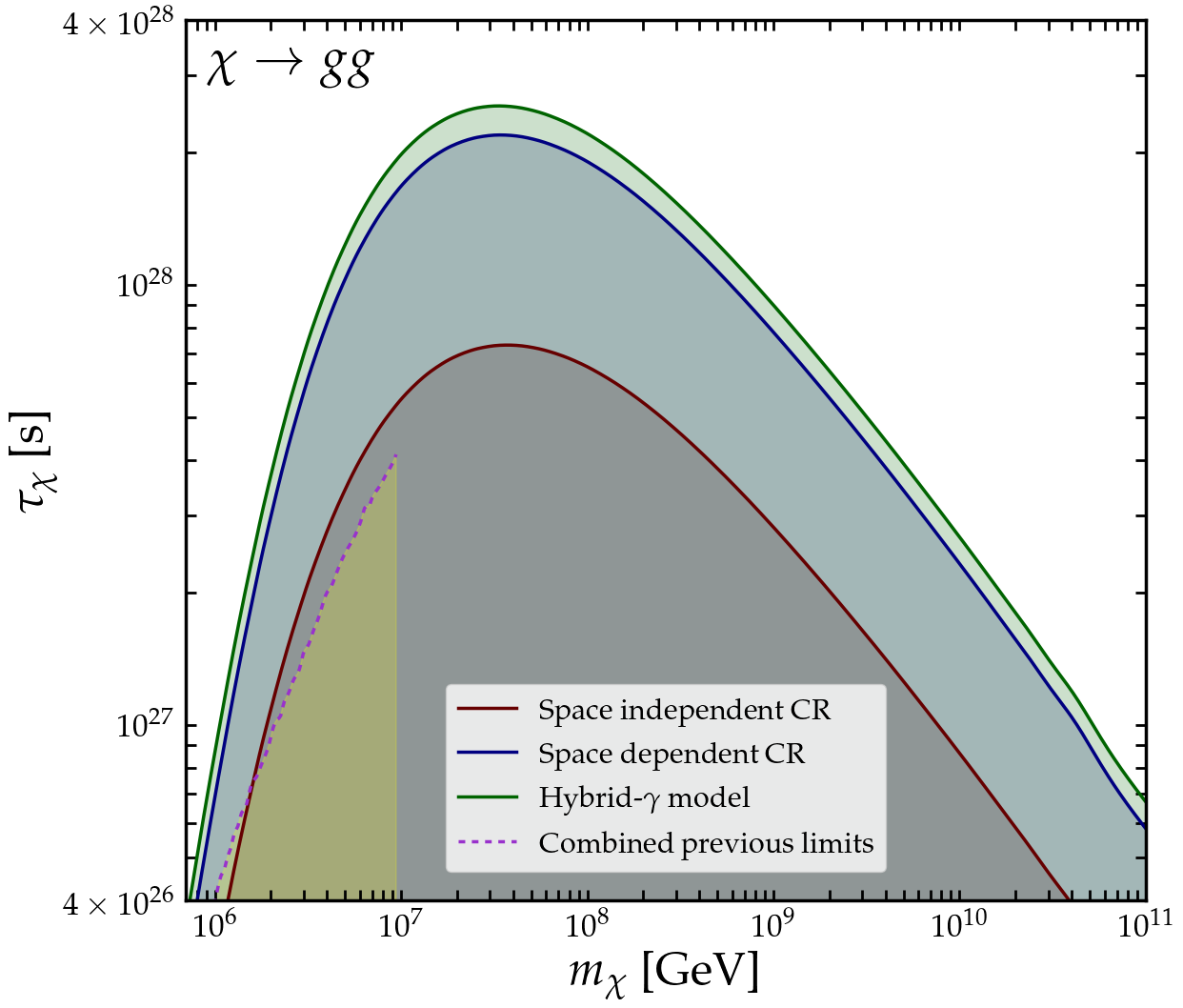}~~\\	
	\includegraphics[angle=0.0,width=0.265\textwidth]{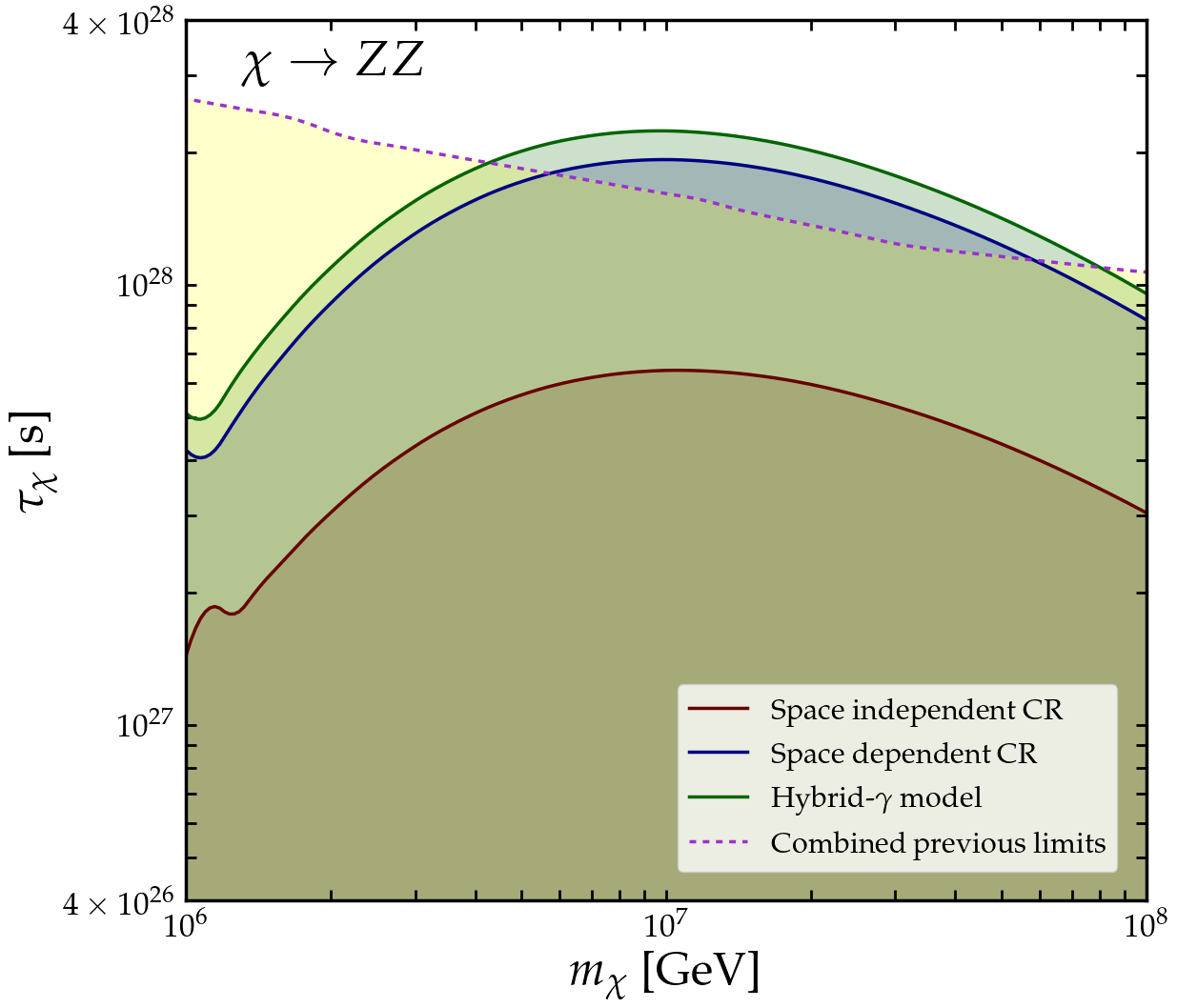}~~
	\includegraphics[angle=0.0,width=0.265\textwidth]{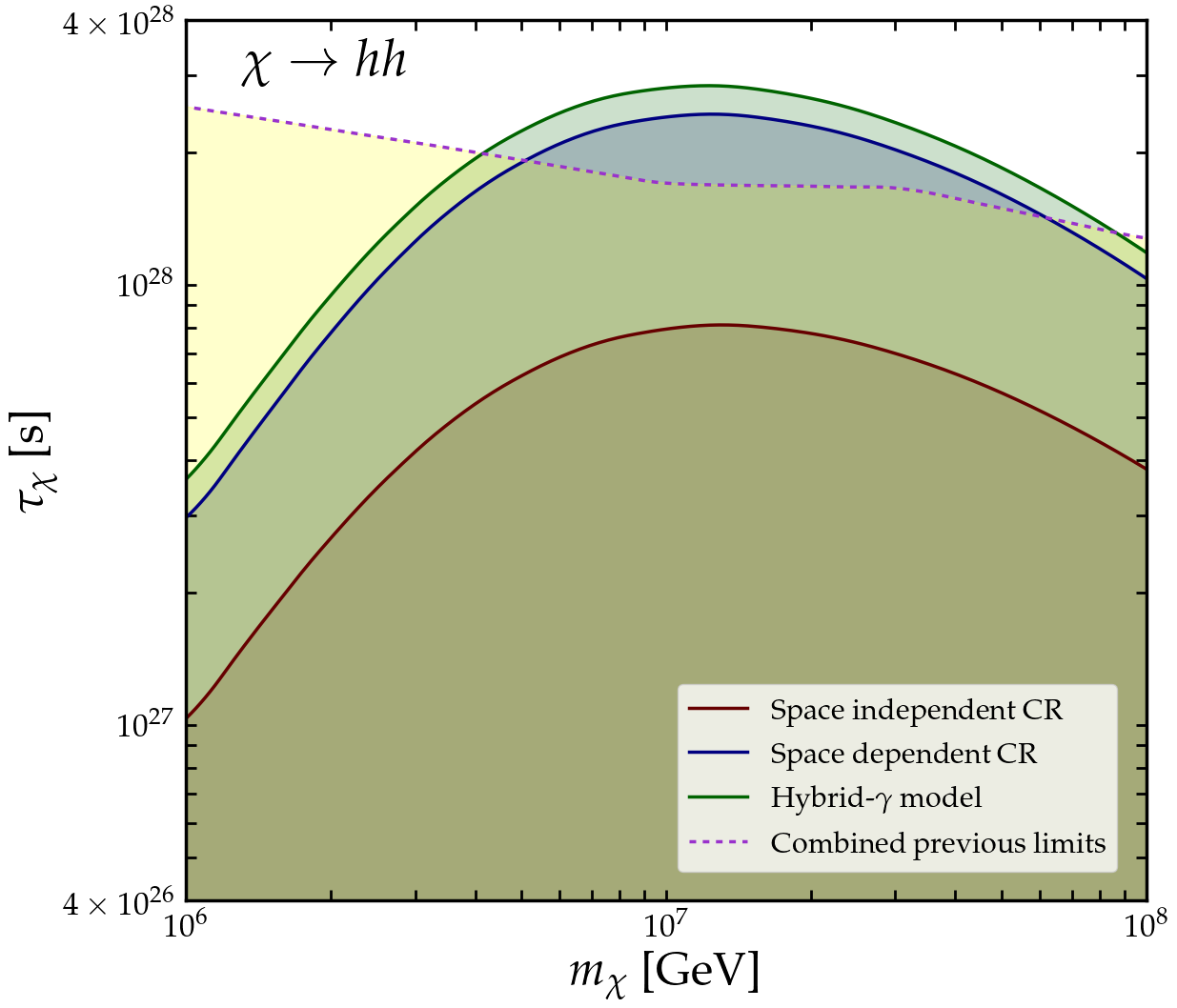}~~
	\includegraphics[angle=0.0,width=0.265\textwidth]{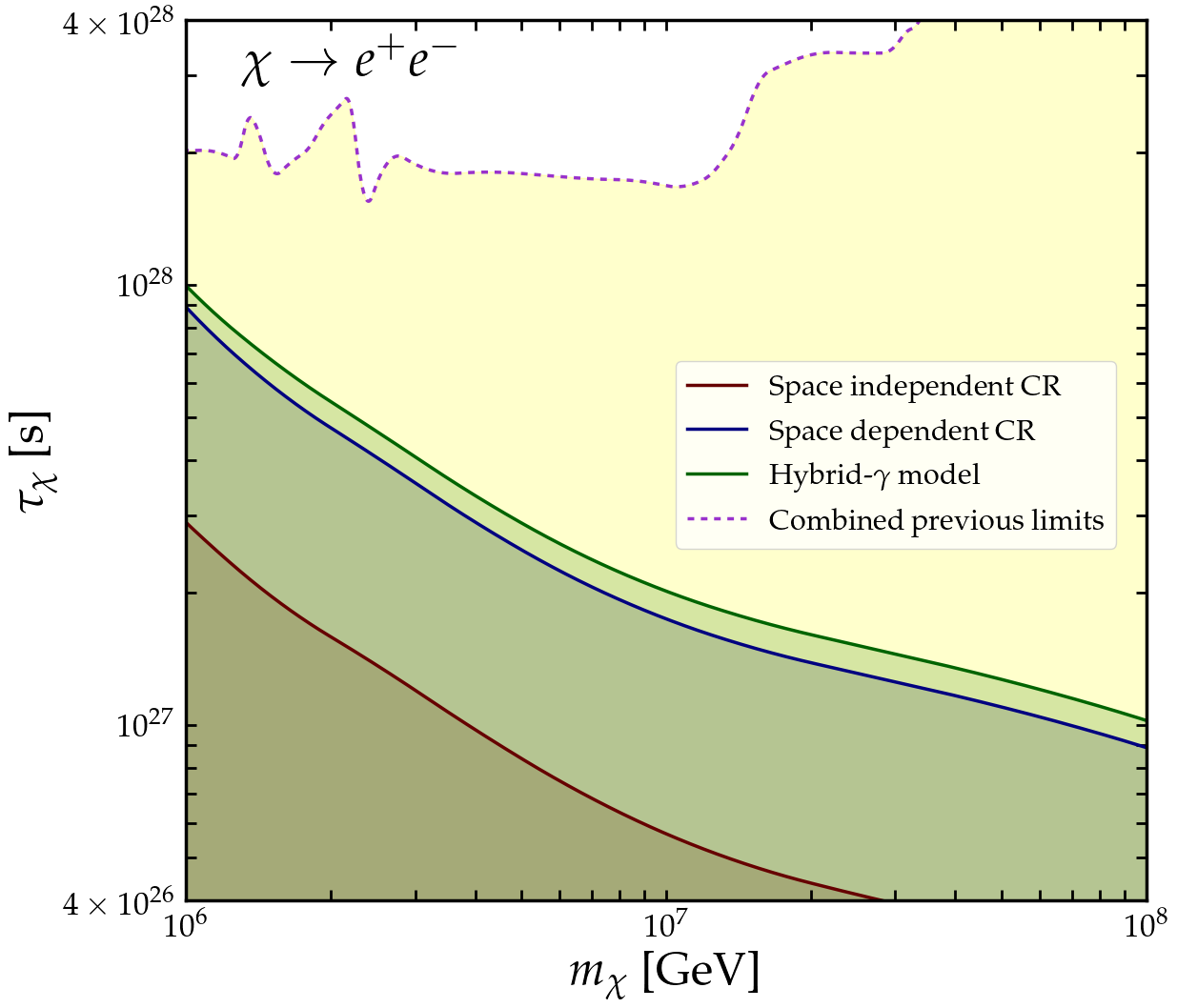}~~\\
    \includegraphics[angle=0.0,width=0.26\textwidth]{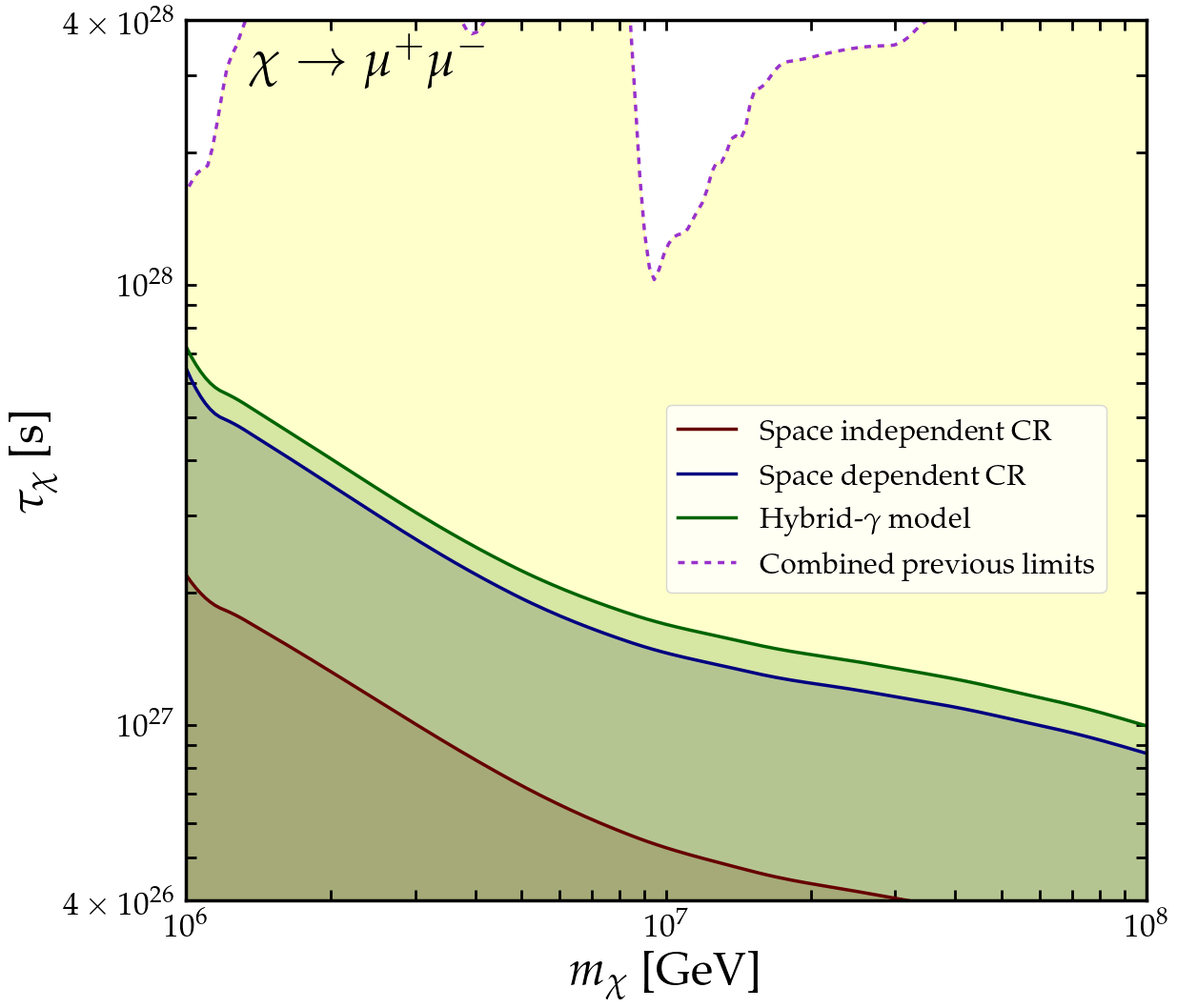}~~	
	\includegraphics[angle=0.0,width=0.26\textwidth]{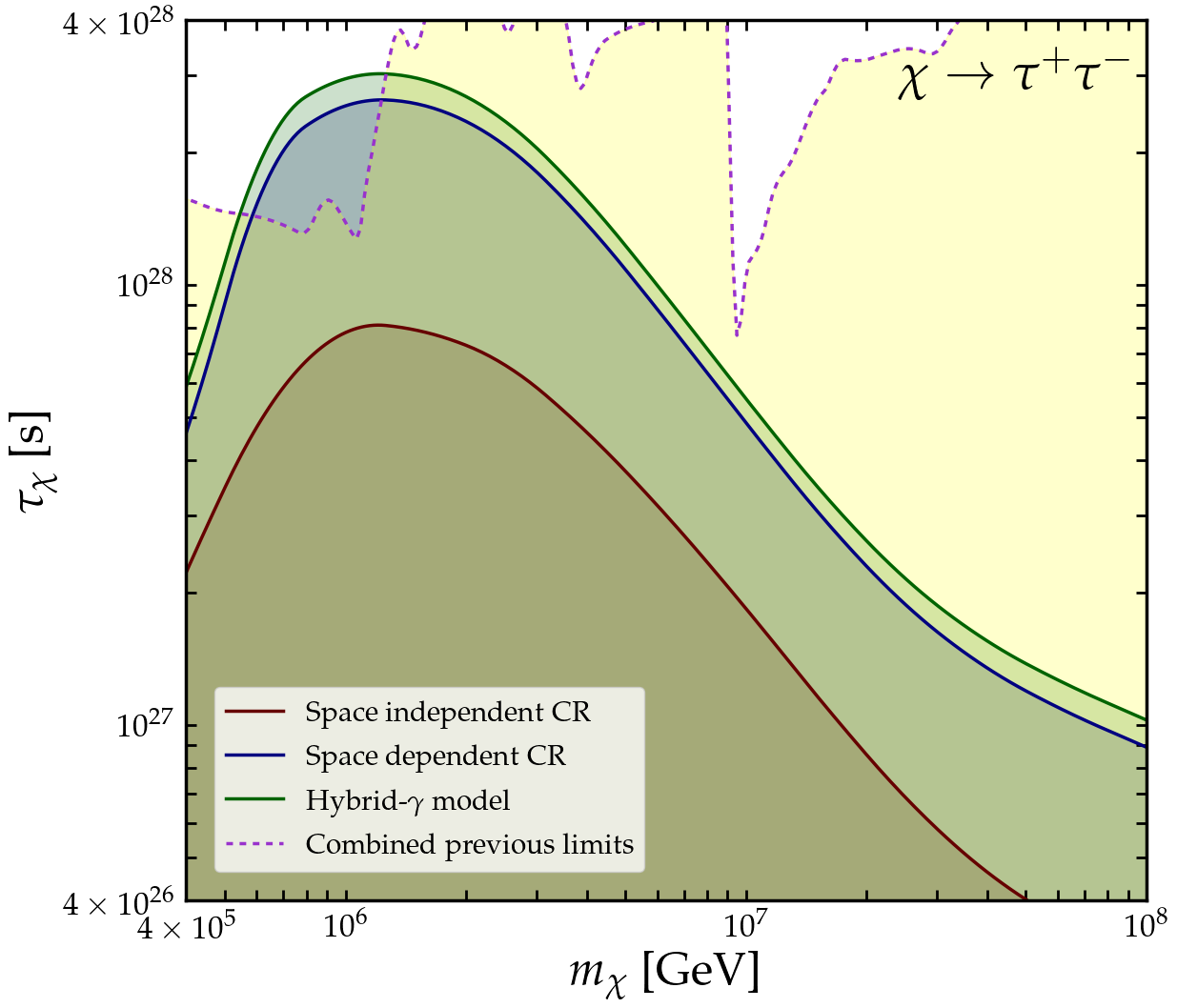}~~
	\includegraphics[angle=0.0,width=0.26\textwidth]{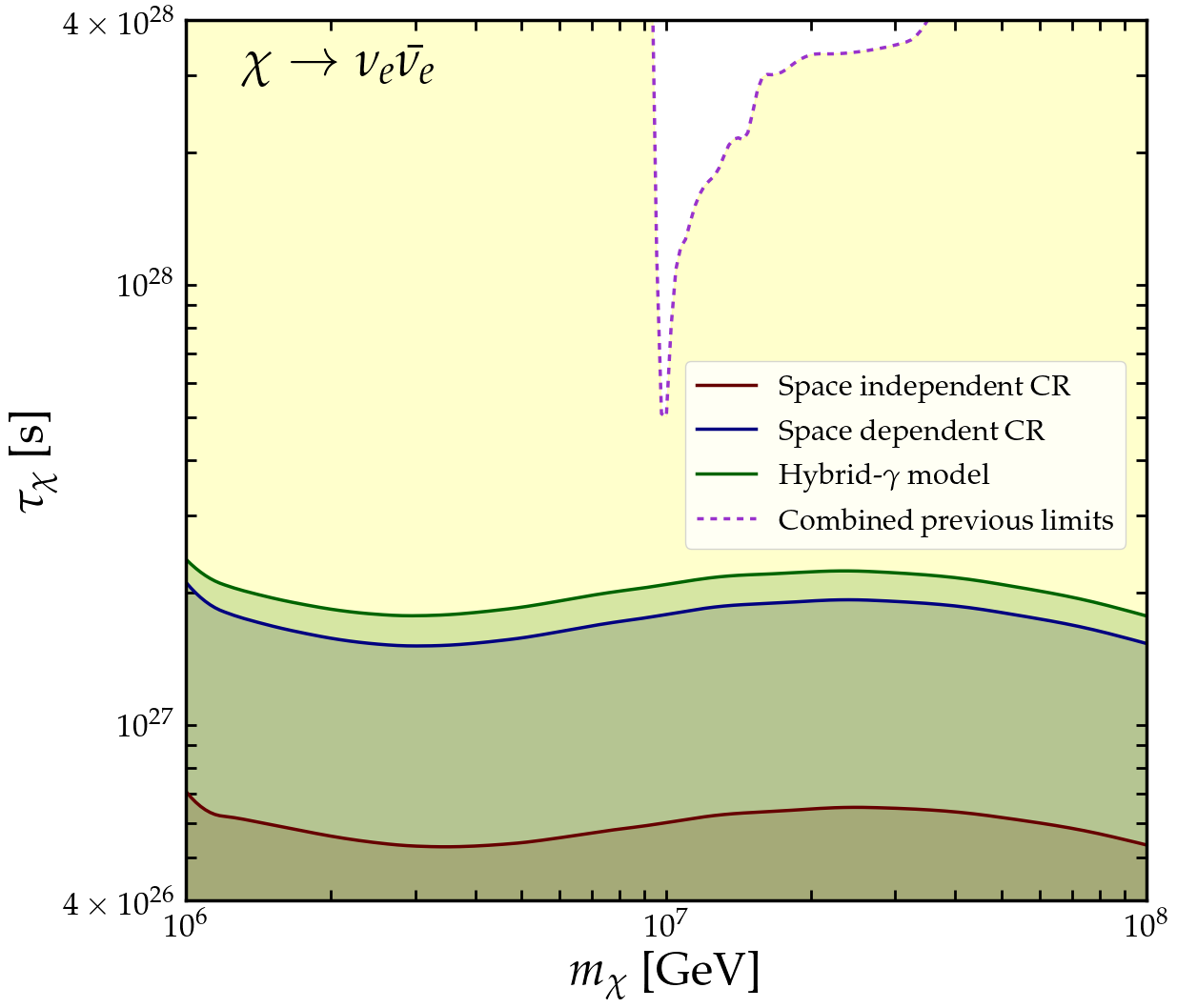}~~\\
	\caption{Upper limits on the DM lifetime as obtained from the Tibet AS$_\gamma$ data assuming various astrophysical production models of sub-PeV gamma-rays. The combined previous constraints from Refs.\,\cite{Cohen:2016uyg, Kachelriess:2018rty, Blanco:2018esa, Bhattacharya:2019ucd, Chianese:2019kyl} are also displayed.}
	\label{fig: Comparison of various final states}
\end{center}	
\end{figure*}

\begin{figure*}[!htbp]
\begin{center}
	\includegraphics[angle=0.0,width=0.45\textwidth]{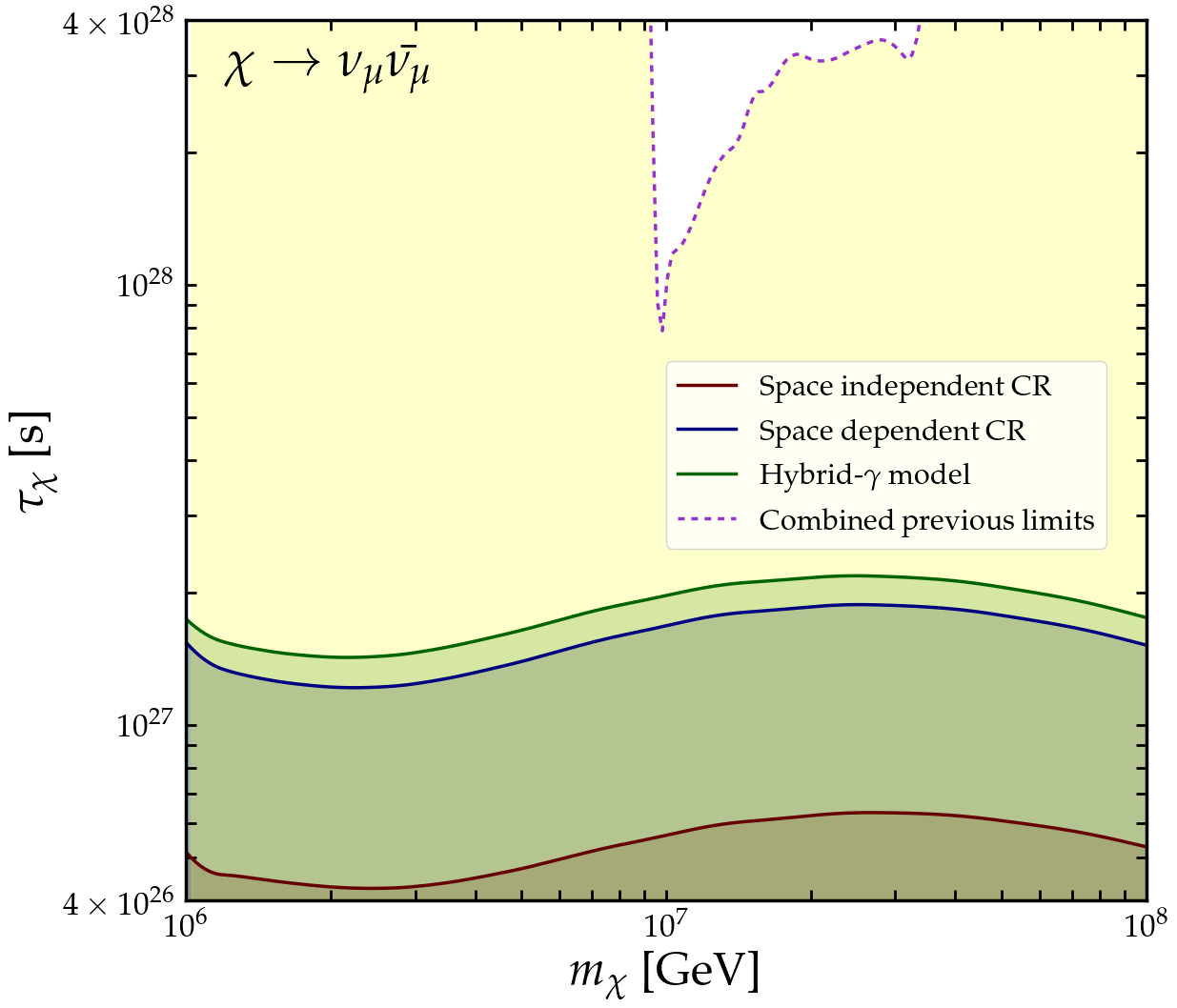}~~
	\includegraphics[angle=0.0,width=0.45\textwidth]{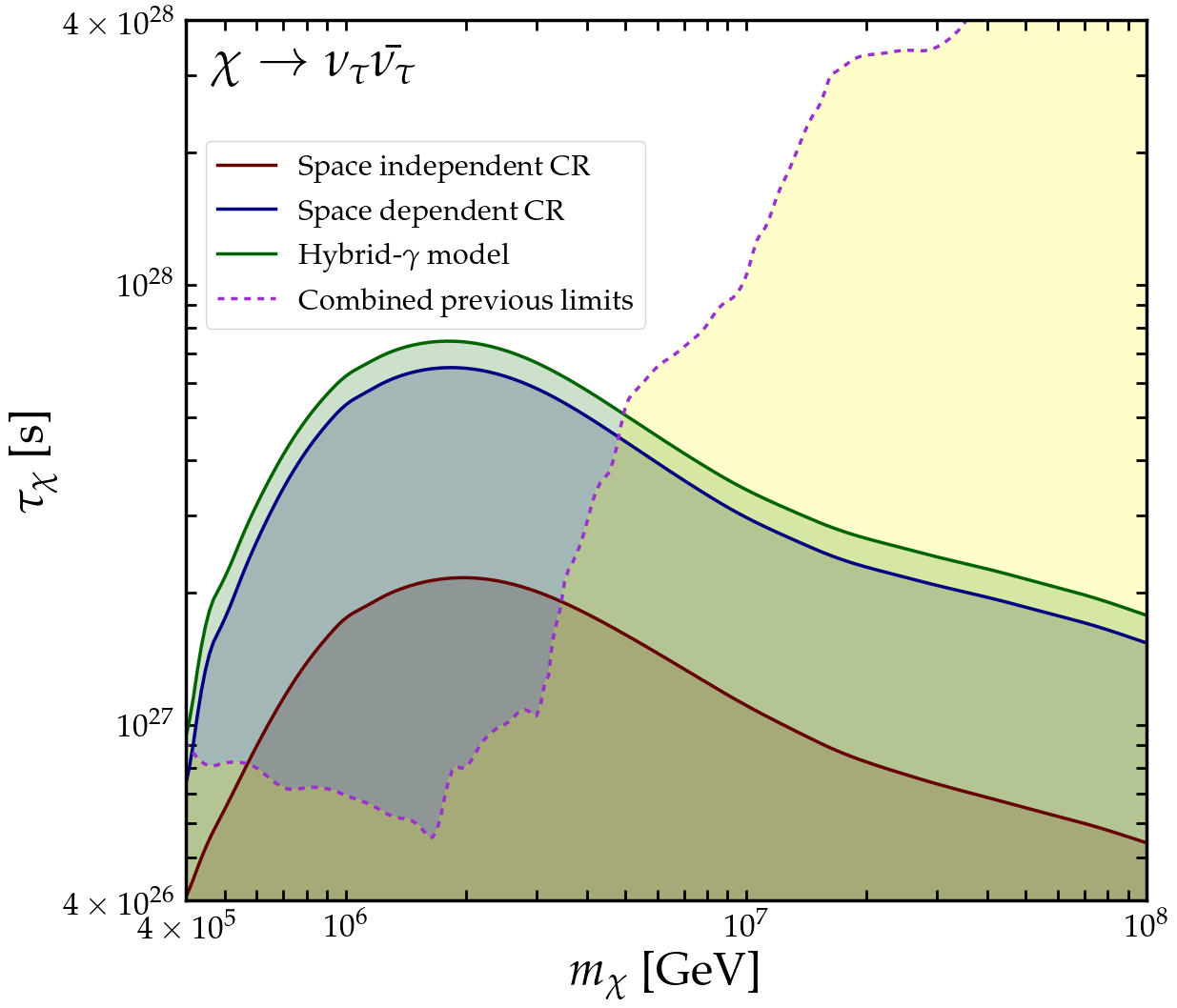}~~\\	
	\caption{Upper limits on the DM lifetime as obtained from the Tibet AS$_\gamma$ data assuming various astrophysical production models of sub-PeV gamma-rays. The combined previous constraints from Refs.\,\cite{Cohen:2016uyg, Kachelriess:2018rty, Blanco:2018esa, Bhattacharya:2019ucd, Chianese:2019kyl} are also displayed.}
	\label{fig: Comparison of various final states2}
\end{center}	
\end{figure*}

\end{document}